\documentclass[10pt, twocolumn]{extarticle}
\usepackage{graphicx}
\usepackage{hyperref}
\usepackage{amssymb}
\usepackage{amsmath}
\usepackage{amsfonts}
\usepackage{parskip}
\usepackage{titling}
\usepackage[table]{xcolor}
\usepackage[square,sort&compress,comma,numbers]{natbib}
\usepackage{multirow}
\usepackage{empheq}

\usepackage{titlesec}
\usepackage{caption}
\titleformat{\section}{\large\bf}{\thesection}{1em}{}
\titleformat{\subsection}{\bf}{\thesubsection}{1em}{}
\titleformat{\subsubsection}{\it}{\thesubsubsection}{1em}{}

\pretitle{\Large\bf\vskip 0.3em} \posttitle{\par\vskip 0.6em} \preauthor{\large} \postauthor{\par\vskip -0.0em}
\predate{\large} \postdate{\par\vskip 0.1em \rule{\linewidth}{0.5pt}} 

\newcommand{\beq}{\begin{equation}}
\newcommand{\eeq}{\end{equation}}
\newcommand{\bea}{\begin{eqnarray}}
\newcommand{\eea}{\end{eqnarray}}

\newcommand{\comment}[1]{}
\renewcommand{\d}{{\rm d}}

\begin{document}
\captionsetup{font=small}

\thanksmarkseries{arabic}

\title{Stability of Self-Configuring Large Multiport Interferometers}
\author{Ryan Hamerly$^{1,2}$, Saumil Bandyopadhyay$^1$, and Dirk Englund$^1$}

\date{August 5, 2022}
\maketitle


{\bf\small Abstract---Realistic multiport interferometers (beamsplitter meshes) are sensitive to component imperfections, and this sensitivity increases with size.  Self-configuration techniques can be employed to correct these imperfections, but not all techniques are equal.  This paper highlights the importance of algorithmic stability in self-configuration.  Na\"ive approaches based on sequentially setting matrix elements are unstable and perform poorly for large meshes, while techniques based on power ratios perform well in all cases, even in the presence of large errors.  Based on this insight, we propose a self-configuration scheme for triangular meshes that requires only external detectors and works without prior knowledge of the component imperfections.  This scheme extends to the rectangular mesh by adding a single array of detectors along the diagonal.}

\begin{flushleft}
\small
$^{1}$      \textit{Research Laboratory of Electronics, MIT, 50 Vassar Street, Cambridge, MA 02139, USA} \\
$^{2}$      \textit{NTT Research Inc., Physics and Informatics Laboratories, 940 Stewart Drive, Sunnyvale, CA 94085, USA}
\end{flushleft}

\rule{\linewidth}{0.5pt}

\section{Introduction}

Photonic technologies increasingly rely on programmable and reconfigurable circuits.  A central component in such circuits is the universal multiport interferometer: an optical device with $N > 2$ inputs and outputs, whose linear input-output relation (transfer matrix) is set by the user.  Such interferometers are indispensable in applications ranging from linear optical quantum computing \cite{Carolan2015, Zhong2020} and RF photonics \cite{Marpaung2013, Zhuang2015} to signal processing \cite{Perez2017, Capmany2012} and machine learning acceleration \cite{Shen2017, Tait2017, Hamerly2019, Shastri2021}, and will play an important role in proposed photonic field-programmable gate arrays \cite{Perez2017, Bogaerts2020}.  Size (i.e.\ number of ports) is an important figure of merit for all of these applications, and scaling up multiport interferometers is an active field in research.  Recent advances in silicon photonics are promising, allowing the scale-up from small proof-of-concept designs to large (and therefore technologically useful) systems \cite{Zhong2020, Harris2020, Ramey2020}.  

Component imperfections are a major challenge to scaling the size of multiport interferometers.  This is because all large devices are based on dense meshes of tunable beamsplitters, whose circuit depth grows with size.  Most non-recirculating designs are variants of the triangular \textsc{Reck} \cite{Reck1994} or rectangular \textsc{Clements} \cite{Clements2016} beamsplitter mesh, both of which encode an $N\times N$ unitary transfer matrix into a compact mesh of programmable Mach-Zehnder interferometers (MZIs).  These circuits have $O(N)$ depth, meaning that component errors cascade as light propagates down the mesh.  The upshot is that scaling in {\it size} must be accompanied by scaling in {\it precision} to preserve the accuracy of the input-output map.  This challenge is most acute for optical machine learning applications \cite{Shastri2021, Shen2017}, which rely on very large mesh sizes for performance \cite{Harris2020, Ramey2020}, where fabrication errors from even state-of-the-art technology are predicted to significantly degrade ONN accuracy in hardware \cite{Fang2019}.

Several self-configuration techniques can suppress the effect of component imprecisions.  For machine learning applications, the MZI phase shifts can be learned by in situ training \cite{Hughes2018}, but this requires extra hardware (inline power detectors \cite{Grillanda2014}) and the learned weights are specific to the given device.  Alternatively, if the chip has been pre-calibrated so the imperfections are known, global optimization can be used to find the phase shifts offline \cite{Burgwal2017, Pai2019}; however, this approach is time-consuming and requires that the hardware imperfections be known to high accuracy.  MZI errors can also be eliminated by pairing MZIs, though this doubles the loss and chip area \cite{Miller2015}.  Finally, for triangular meshes, the MZIs of each diagonal can be configured sequentially \cite{Miller2013a, Miller2013b, Miller2017}.  This approach, however, also requires $O(N^2)$ inline power detectors (or pre-calibrated MZIs that can be configured to a perfect ``bar'' or ``cross'' state).  In short, all configuration schemes to date rely on either (i) additional hardware complexity, i.e.\ inline detectors or MZI pairing, or (ii) accurate pre-calibration of the mesh's component errors.

In this article, we analyze self-configuration algorithms that require only external detectors and do not rely on prior calibration of the MZI mesh.  Not all algorithms are created equal, and algorithmic stability distinguishes good algorithms from bad ones: for example, a straightforward approach based on sequentially matching matrix elements works in principle, but performs poorly in the presence of large errors.  Based on this insight, we propose an algorithm based on orthogonality and power ratios that performs well in all cases, even in the presence of large errors.  This scheme is directly applicable to triangular meshes, but can also be extended to a rectangular mesh with the addition of a single array of inline power detectors along the diagonal.

This paper is organized as follows: In Sec.~\ref{sec:stats}, we introduce the \textsc{Reck} and \textsc{Clements} meshes, analyze their statistical properties in the presence of errors, and derive analytic estimates for the improvements possible by self-configuration.  Sec.~\ref{sec:alg} covers the theory of self-configuring algorithms and introduces our proposed methods in the context of the \textsc{Reck} scheme.  Sec.~\ref{sec:perf} analyzes the accuracy of self-configuration in the presence of component errors and highlights the importance of algorithmic stability.  Finally, Sec.~\ref{sec:clem} extends our method to the rectangular \textsc{Clements} scheme by splitting the rectangle with a diagonal of internal monitors.

\section{Statistics of Imperfect Meshes}
\label{sec:stats}

\begin{figure}[b!]
\begin{center}
\includegraphics[width=1.00\columnwidth]{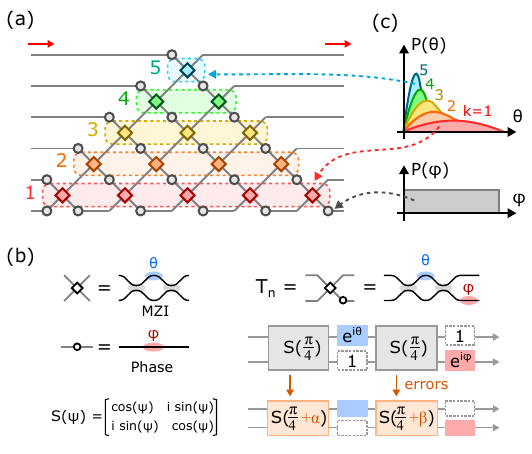}  
\caption{(a) Triangular (\textsc{Reck}) universal multiport interferometer \cite{Reck1994}.  (b) Constituent components and $2\times 2$ unit cell $T_n$, illustrating the effect of errors on the transfer matrix.  (c) Distribution of splitting angles $\theta$ and phase shifts $\phi$ over the Haar measure, Eq.~(\ref{eq:pph}-\ref{eq:pth}).}
\label{fig:f1}
\end{center}
\end{figure}

The most common multiport interferometer designs are the \textsc{Reck} triangle \cite{Reck1994} and the \textsc{Clements} rectangle \cite{Clements2016}.  In both cases, the circuit can be laid out on a regular grid of $2\times 2$ elements (Fig.~\ref{fig:f1}(a)) without any waveguide crossings, a major advantage compared to competing designs \cite{Lopez2019, Fldzhyan2020, Saygin2020, Polcari2018}. 
The input-output matrix of the mesh is accordingly a product:
\beq
	U = \Bigl(\sum_n T_n\Bigr) D \label{eq:tprod}
\eeq  
where $D$ is a 
phase mask and the $T_n$ are $2\times 2$ block matrices representing a phase shifter cascaded into an MZI crossing (Fig.~\ref{fig:f1}(b)):
\bea
	T_n & \!=\! & 
	\underbrace{\begin{bmatrix} 1 & 0 \\ 0 & e^{i\phi_n} \end{bmatrix}}_{P_2(\phi_n)} 
	\underbrace{\begin{bmatrix} \cos(\tfrac\pi4 + \beta_n) & i\sin(\tfrac\pi4 + \beta_n) \\ i\sin(\tfrac\pi4 + \beta_n) & \cos(\tfrac\pi4 + \beta_n) \end{bmatrix}}_{S(\tfrac\pi4 + \beta_n)} \nonumber \\
	& & \times 
	\underbrace{\begin{bmatrix} e^{i\theta_n} & 0 \\ 0 & 1 \end{bmatrix}}_{P_1(\theta_n)}
	\underbrace{\begin{bmatrix} \cos(\tfrac\pi4 + \alpha_n) & i\sin(\tfrac\pi4 + \alpha_n) \\ i\sin(\tfrac\pi4 + \alpha_n) & \cos(\tfrac\pi4 + \alpha_n) \end{bmatrix}}_{S(\tfrac\pi4 + \alpha_n)}
	 \label{eq:tprod2}
\eea
Here $(\theta_n, \phi_n)$ are the phases programmed by the user, e.g.\ through thermo-optic \cite{Harris2014}, or MEMS \cite{Ramey2020} phase shifters, while $(\tfrac\pi4 + \alpha_n, \tfrac\pi4 + \beta_n)$ are the coupler angles, a property of the circuit and its imperfections ($\alpha_n, \beta_n$).  These angles are $\pi/4$ in an ideal MZI, which enables perfect contrast on each MZI output.  In such a device, the phase shifts can be found by a procedure that diagonalizes $U$ with a sequence of $2\times 2$ rotations \cite{Reck1994, Clements2016}. 

(An equally valid convention is to place the phase mask at the end, $U = D \prod_n T_n$, and put the phase shifter $\phi_n$ at the beginning of the unit cell: $T_n = S(\tfrac\pi4+\beta_n) P_1(\theta_n) S(\tfrac\pi4+\alpha_n) P_1(\phi_n)$; however, the algorithm described in Sec.~\ref{sec:alg} is easiest to adapt to the convention in Eqs.~(\ref{eq:tprod}-\ref{eq:tprod2})).

\subsection{Component Errors}

Component errors (deviations from design) will perturb the input-output matrix.  We are primarily interested in the magnitude of this perturbation $\Delta U$, quantified by the Frobenius norm $\lVert \Delta U \rVert^2 = \sum_{ij} |\Delta U_{ij}|^2$ and normalized to define an error measure:
\beq
	\mathcal{E} = \frac{1}{\sqrt{N}} \langle \lVert \Delta U \rVert \rangle_{\rm rms} \in [0, 2] \label{eq:err}
\eeq
This metric can be interpreted as an average relative error per entry in the matrix $U$; for small $\mathcal{E}$, the quantity $(1 - \mathcal{E})$ plays a role similar to the fidelity of a quantum operation.
To first order, the effect of component errors is linear:
\beq
	\Delta U = \sum_n \Bigl[\Bigl(\frac{\partial U}{\partial\alpha_n}\Bigr)\alpha_n + \Bigl(\frac{\partial U}{\partial\beta_n}\Bigr)\beta_n \Bigr] \label{eq:du}
\eeq
Applying Eqs.~(\ref{eq:tprod}-\ref{eq:tprod2}), we find $\partial U/\partial \alpha_n = U_{\rm pre} S'(\pi/4) U_{\rm post}$ (and likewise for $\partial U/\partial\beta_n$).  Since we are interested here in the magnitude $\lVert\Delta U\rVert$ and matrices $(U_{\rm pre}, U_{\rm post})$ are unitary, it follows that:
\beq
	\Bigl\lVert \frac{\partial U}{\partial\alpha_n}\Bigr\rVert = \Bigl\lVert \frac{\partial U}{\partial\beta_n}\Bigr\rVert = \lVert S'(\pi/4) \rVert = \sqrt{2}
\eeq
Thus, the mesh is equally sensitive to all beamsplitters, irrespective of geometry.

At this point it becomes necessary to introduce an error model, since perturbations from nearby crossings may lead to correlated errors in $U$.  While real imperfections are correlated, this adds significant complexity to the math that obfuscates the insights.  Moreover, while correlations may affect the error measure of a particular matrix, when considering ensembles of matrices, they are expected to average to zero (see Appendix~\ref{sec:appcorr}).  Therefore, for the remainder of this paper, we assume an uncorrelated error model where $\langle \alpha_n \rangle_{\rm rms} = \langle \beta_n \rangle_{\rm rms} = \sigma$ for a given error amplitude $\sigma$.

Under the uncorrelated model, the error terms in Eq.~(\ref{eq:du}) add in quadrature over the $N(N-1)$ couplers to give:
\beq
	\mathcal{E} = \sqrt{2(N-1)}\,\sigma \sim (2N)^{1/2}\sigma  \label{eq:e0}
\eeq
Since the depth of the circuit is $O(N)$, and independent errors in each layer add in quadrature, it is not surprising that $\mathcal{E} \propto N^{1/2}$.  Component precision therefore must increase as meshes are scaled up in order to maintain a desired matrix accuracy.

\subsection{Error Correction}
\label{sec:errcorr}

As mentioned earlier, if the imperfections $(\alpha_n, \beta_n)$ are known, correction schemes can be used to program a unitary to, in most cases, better accuracy than Eq.~(\ref{eq:e0}).  Recently, we presented a straightforward ``local'' scheme to correct for imperfections at each MZI separately \cite{SaumilPaper} (see also Ref.~\cite{Kumar2021}).  The method is based on the observation that $2\times 2$ unitaries with the same power splitting ratio are equivalent up to external phase shifts.  In the presence of imperfections, the MZI does not achieve perfect contrast in both output ports.  The range of splitting angles is truncated to:
\beq
	2|\alpha + \beta| \leq \theta \leq \pi - 2|\alpha - \beta| \label{eq:thetabd}
\eeq
Provided Eq.~(\ref{eq:thetabd}) is satisfied, a perfect MZI $T(\theta_n, \phi_n)$ can be replaced by an imperfect MZI with external phase shifts:
\begin{align}
	& \exists\  \theta'_n, \phi'_n, \chi'_n, \psi'_n: \nonumber \\
	& T(\theta_n, \phi_n) 
	= T(\theta'_n, \phi'_n | \alpha_n, \beta_n) 
	\begin{bmatrix} e^{i\chi'_n} & 0 \\ 0 & e^{i\psi'_n} \end{bmatrix}
\end{align}
(The extra phase shifts can be absorbed into the neighboring MZIs so that the number of physical phase shifters on the mesh does not increase.)  Provided that Eq.~(\ref{eq:thetabd}) is satisfied for all MZIs in the ideal \textsc{Reck} / \textsc{Clements} decomposition, this procedure in Ref.~\cite{SaumilPaper} leads to a perfect representation of the matrix.  The fraction of unitary matrices realizable by this imperfect mesh is called the {\it coverage}, $\text{cov}(N, \sigma)$.  If some MZIs do not satisfy Eq.~(\ref{eq:thetabd}), we pick the closest possible $\hat{\theta}_n \in \{\theta_{\rm min}, \theta_{\rm max}\}$, which leads to an error in the matrix:
\bea
	\lVert \Delta U \rVert_{\rm MZI} & = & \biggl\lVert \begin{bmatrix} \cos(\theta_n/2) & i\sin(\theta_n/2) \\ i\sin(\theta_n/2) & \cos(\theta_n/2) \end{bmatrix} \nonumber \\
	& &  \ \ - \begin{bmatrix} \cos(\hat{\theta}_n/2) & i\sin(\hat{\theta}_n/2) \\ i\sin(\hat{\theta}_n/2) & \cos(\hat{\theta}_n/2) \end{bmatrix} \biggr\rVert \nonumber \\
	& = & 2^{-1/2} |\theta_n - \hat{\theta}_n| + O\bigl((\theta_n - \hat{\theta}_n)^3\bigr) \label{eq:dumzi}
\eea

\begin{figure}[b!]
\begin{center}
\includegraphics[width=1.00\columnwidth]{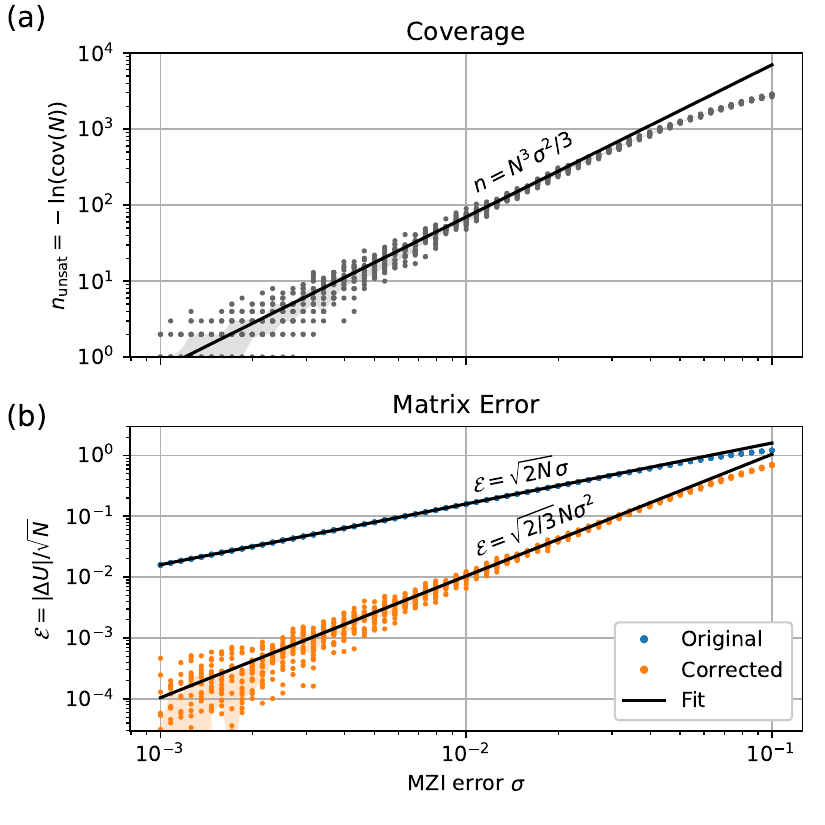}  
\caption{Realization of Haar-random unitaries with imperfect meshes, $N = 128$.  (a) Number of unsatisfied MZIs, which is related to the coverage $n_{\rm unsat} = -\ln(\text{cov}(N))$, Eq.~(\ref{eq:cov}).  (b) Matrix error $\mathcal{E} = \lVert \Delta U\rVert / \sqrt{N}$ with and without correction, demonstrating the accuracy enhancement of the ``local'' correction algorithm.}
\label{fig:f2}
\end{center}
\end{figure}

Not all unitaries are equally easy to express on an MZI mesh.  For this reason, when analyzing the efficiency of an error correction scheme, one must specify the probability distribution of $U$.  Here, we consider random unitaries under the Haar measure, a distribution that samples uniformly from the space of unitary matrices \cite{Haar1933, Tung1985}.  Under this measure, the phase shifts $\theta_n \in [0, \pi)$ and $\phi_n \in [0, 2\pi)$ are uncorrelated and distributed according to \cite{Russell2017}:  
\bea
	P(\phi) & = & \frac{1}{2\pi}\ \ \ \text{(uniform over 
	 $[0, 2\pi)$
	 )} \label{eq:pph} \\
	P(\theta|k) & = & k \sin(\theta/2)\cos(\theta/2)^{2k-1} \label{eq:pth}
\eea
where $k$ is the row index of the \textsc{Reck} mesh, starting from $k = 1$ at the bottom (Fig.~\ref{fig:f1}(a)).  The phase shifts of the \textsc{Clements} mesh follow the same distribution with a different layout of crossings \cite{Russell2017}.  The distribution Eq.~(\ref{eq:pth}) clusters tightly around $\theta = 0$ for MZIs with large $k$ \cite{Russell2017, Burgwal2017, Pai2019} (Fig.~\ref{fig:f1}(c)).  Therefore, coverage and accuracy in large meshes is primarily limited by the small $\theta$ values.  $P(\theta|k)$ can be linearized for small $\theta$, so the probability of a single MZI breaking the bound Eq.~(\ref{eq:thetabd}) is approximately
\beq
	p_{\rm unsat}(k) = \int_0^{2|\alpha+\beta|} P(\theta|k)\d\theta \approx k (\alpha+\beta)^2
\eeq
For an $N\times N$ unitary, there are $(N-k)$ MZIs of rank k.  The coverage is equal to the probability, under the Haar measure, that all $\theta_n$ are realizable.  Since the $\theta_n$ are uncorrelated, this is a product of the probabilities for each MZI:
\begin{align}
	& \text{cov}(N) = \prod_k \bigl(1 - p_{\rm unsat}(k)\bigr)^{N-k} \nonumber \\
	& \ \ \approx \exp\Bigl[-\underbrace{\sum_k k(N-k)\langle (\alpha+\beta)^2}_{n_{\rm unsat}}\rangle\Bigr]
	\approx e^{-N^3\sigma^2/3} \label{eq:cov}
\end{align}
This is vanishingly small for large MZI meshes: for example, taking a reasonable value of $\sigma = 0.02$, even a $32\times 32$ mesh has a coverage around 1\%.  In general, the number $n_{\rm unsat}$ of ``unsatisfiable'' MZIs that break condition (\ref{eq:thetabd}) increases rapidly with error and problem size (Fig~\ref{fig:f2}(a)).

Even if most unitaries cannot be realized exactly, they can be approximated to much better accuracy than the uncorrected result Eq.~(\ref{eq:e0}).  Each MZI with an unrealizable $\theta_n$ will lead to a matrix error per Eq.~(\ref{eq:dumzi}).  Over the Haar measure, the average error induced by a particular MZI is thus:
\begin{align}
	& \langle \lVert \Delta U \rVert^2 \rangle_{{\rm MZI},k} = \int{P(\theta|k) \lVert \Delta U(\theta) \rVert^2 \d\theta} \nonumber \\
	& \approx \int_0^{2|\alpha+\beta|}{\frac{k\theta}{2} \frac{(2|\alpha+\beta| - \theta)^2}{2}\d\theta}
	= \frac{k}{3} (\alpha+\beta)^4 \label{eq:dumzi2}
\end{align}
Assuming the errors are uncorrelated (see Appendix~\ref{sec:appcorr} for the correlated case), they add in quadrature: $\langle \lVert \Delta U \rVert^2 \rangle = \sum_k (N-k) \langle \lVert \Delta U \rVert^2 \rangle_{k} \approx \tfrac{1}{18} N^3 \langle (\alpha+\beta)^4\rangle = \frac{2}{3} N^3 \sigma^4$.  Following Eq.~(\ref{eq:err}), the corrected normalized error is therefore:
\beq
	\mathcal{E}_{\rm corr} = \sqrt{2/3}\,N \sigma^2 \label{eq:ecorr}
\eeq
This is plotted in Fig.~\ref{fig:f2}(b).  Recall from Eq.~(\ref{eq:e0}) that the uncorrected error scales as $\mathcal{E} \propto \sqrt{N}\,\sigma$.  This means that $\mathcal{E}_{\rm corr} \propto \mathcal{E}^2$, i.e.\ correction allows for an effective ``squaring'' of the error.  If the errors are very large to begin with, error correction will not provide much benefit.  However, for most fabricated circuits the uncorrected $\mathcal{E}$ is reasonably small (though not small enough for many applications), and error correction can give a significant boost in accuracy.

\section{Self-Configuring Algorithms}
\label{sec:alg}

In many circumstances, the correction procedure in Sec.~\ref{sec:errcorr} cannot be applied because the errors in an MZI mesh are known to sufficient accuracy.  Nevertheless, for triangular meshes, ``progressive'' self-configuration strategies can still be used.  As noted earlier, existing strategies rely on inline photodetectors or pre-calibration, which limits their usefulness in many systems \cite{Miller2013a, Miller2013b, Miller2017, Pai2020}.  This section introduces two schemes that do not rely on these assumptions and can be run on uncalibrated hardware with only external detectors: a simple ``direct'' method based on sequentially setting matrix elements and a ``ratio'' method based on setting power ratios.  While both schemes work in principle, only the ratio method is robust in the presence of large errors.  This distinction highlights the importance of algorithmic stability when configuring large multiport interferometers.

\subsection{Direct Method}

\begin{figure}[b!]
\begin{center}
\includegraphics[width=1.00\columnwidth]{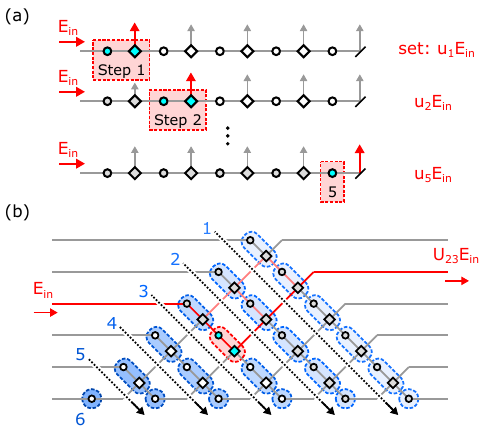}
\caption{Direct method for programming triangular MZI meshes.  (a) Procedure for programming a 1:N splitter, starting at the input MZI and working down to the last output.  (b) Programming the \textsc{Reck} triangle.  This involves starting at the top diagonal and working down, applying the procedure in Fig.~\ref{fig:f3}(a) to each diagonal.}
\label{fig:f3}
\end{center}
\end{figure}

The simplest way to program a triangular mesh is to set the MZI phases one at a time to match the target matrix elements $\hat{U}_{ij}$.  This is easiest to understand by considering first the case of a tunable 1:N splitter and later generalizing to an $N\times N$ unitary.  Fig.~\ref{fig:f3}(a) shows the ``direct'' method for programming a 1:N splitter, which we wish to set to a target splitting vector $\hat{u}$, $\lVert \hat{u} \rVert = 1$.  Given a coherent input $E_{\rm in}$, the procedure consists of $N$ steps, starting from the input of the splitter and working towards the final output.  In the first $N-1$ steps, a pair of phases $(\theta, \phi)$ (corresponding to an MZI and phase shifter, Fig.~\ref{fig:f1}(a)) are tuned to set the $m^{\rm th}$ (complex) output amplitude to $\hat{u}_m E_{\rm in}$.  The two degrees of freedom are sufficient to independently set the real and imaginary components of $\hat{u}_m$.  In the final step, there is only a single degree of freedom (a phase shift $\phi$); however, since $\hat{u}$ has unit norm, at this stage its amplitude is already constrained and only the phase is free; therefore, provided the preceding elements $\hat{u}_m$ are set properly, only single phase shift is needed to set $\hat{u}_N$.

Fig.~\ref{fig:f3}(b) shows the direct method for programming a \textsc{Reck} triangle, which can be divided into $N$ diagonals, each functioning as a tunable one-to-many splitter, with the output of each diagonal fed into the inputs of its upper-right neighbor.  The triangle is configured from the top diagonal to the bottom, working down each diagonal.  Each MZI element $(\theta_{mn}, \phi_{mn})$ (the $n^{\rm th}$ element of the $m^{\rm th}$ diagonal, starting from the top) is configured to set the matrix element $\hat{U}_{nm}$ between output $n$ and input $m$.  In this way, the lower diagonal of $\hat{U}$ is correctly configured -- which, given the unitarity of $\hat{U}$, correctly configures the entire matrix.

The matrix must be triangular in order for this procedure to work.  Triangularity guarantees that tuning steps do not disturb the matrix elements that have already been set, provided that the order in Eq.~\ref{fig:f3}(b) is followed.  Thus, the direct  method cannot configure the \textsc{Clements} matrix, although we show in Sec.~\ref{sec:clem} that \textsc{Clements} can be divided into two triangles, which can be separately configured.

\subsection{Ratio Method}
\label{sec:ratio}

\begin{figure}[b!]
\begin{center}
\includegraphics[width=1.00\columnwidth]{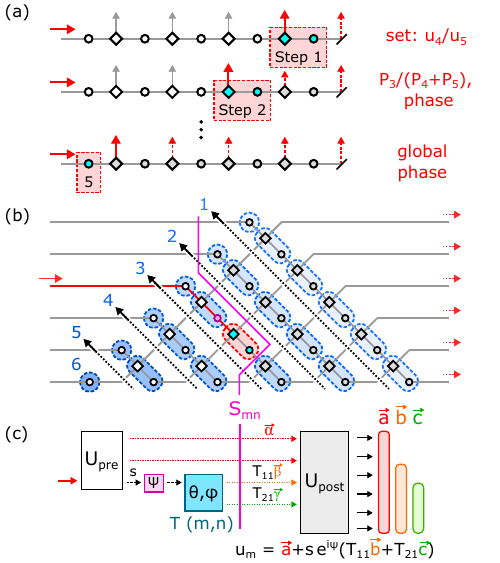}
\caption{Ratio method for configuring MZI meshes.  (a) Programming a 1:N splitter.  (b) MZI order for programming an $N\times N$ \textsc{Reck} triangle.  (c) Output field during configuration of MZI $(m, n)$ decomposed into three components according to the path of the light.}
\label{fig:f4}
\end{center}
\end{figure}

One can also configure the mesh by a method based on power ratios.  As before, it is easiest to describe this method in the case of a 1:N splitter (Fig.~\ref{fig:f4}(a)) and then generalize to the \textsc{Reck} triangle.  In this case, the splitter phases are configured in reverse order to set the {\it power ratios} (and relative phases) of the outputs.  As before, let $\hat{u}$ be the target vector and $\vec{u}$ be the output of the physical splitter.  The configuration steps are as follows:
\begin{itemize}
	\item {\it Step 1}: Set splitter angle $\theta$ to match the power ratio $|u_{N-1}/u_N| = |\hat{u}_{N-1}/\hat{u}_N|$.  Next, set phase shift $\phi$ to match the relative phase $\text{arg}(u_{N-1}/u_N) = \text{arg}(\hat{u}_{N-1}/\hat{u}_N)$ between the last two outputs.
	\item {\it Intermediate Steps}: Here, we configure the phase shifts corresponding to the $n^{\rm th}$ output, $1 \leq n < N$.  These are set to align the partial output vectors: $\vec{u}_{n:N} \parallel \hat{u}_{n:N}$ (here $\vec{a}_{n:N} = [a_n, \ldots, a_N]$ denotes the slice of a vector $\vec{a}$ over a given index set, red outputs in Fig.~\ref{fig:f4}(a)).  The splitter angle $\theta$ is set to match the power ratios $P_n/(P_{n+1} + \ldots + P_N)$, while the phase shift is used to compensate any relative phases.  Overall, this corresponds to maximizing the inner product $\text{max}_{\theta,\phi} \bigl|\langle \vec{u}_{n:N} | \hat{u}_{n:N} \rangle \bigr|$.
	
	Without errors, this is equivalent to matching the amplitude and phase of $u_n/u_{n+1}$ as all downstream ratios have already been configured; however, using all the outputs in the configuration is more robust to errors (especially when $u_{n+1}$ is small).
	\item {\it Step N}: Set the final phase shift to align the phase of $\vec{u}$ with $\hat{u}$.
\end{itemize}

The \textsc{Reck} triangle is configured one diagonal at a time in the order shown in Fig.~\ref{fig:f4}(b).  Here, we have indexed the each MZI $(m, n)$ according to its diagonal ($m$) and position relative to the triangle base ($n$).  When configuring an MZI along the $m^{\rm th}$ diagonal, light enters port $m$ so that only the top port of the MZI is excited.  All MZIs downstream from $(m, n)$ have been tuned, while upstream MZIs are untuned, and a spacelike separator $S_{mn}$ (purple line in figure) divides the configured and unconfigured parts of the mesh.  We can write the unitary of this circuit as:
\beq
	U = U_{\rm post} T U_{\rm pre}
\eeq
where $T = \bigl[[T_{11}, T_{12}], [T_{21}, T_{22}]\bigr]$ is the MZI transfer function, which depends on the phases $(\theta_{mn}, \phi_{mn})$.  The output field is a sum of 3 contributions (Fig.~\ref{fig:f4}(c)):
\begin{enumerate}
	\item Light that bypasses the MZI $(m,n)$.  At surface $S_{mn}$, this is denoted by $\vec{\alpha}$, and at the output it is $\vec{a} = U_{\rm post}\vec{\alpha}$.
	\item Light that enters $(m,n)$ and leaves through its top port.  The input light has an unknown amplitude $s\,e^{i\psi}$ ($\psi$ set by the upstream phase shifter, purple), but only relative amplitudes matter when configuring $(m,n)$.  The output to the top port is $s\,e^{i\psi} T_{11}$.  At surface $S_{mn}$, the field is denoted by the vector $s\,e^{i\psi} T_{11} \vec{\beta}$, where $\vec{\beta} = \hat{e}_{N-n}$ is the one-hot vector for waveguide $(N-n)$ (the top output of $(m, n)$).  Thus the output field is $s\,e^{i\psi} T_{11}\vec{b}$, where $b = U_{\rm post}\vec{\beta}$.
	\item Light that enters $(m,n)$ and leaves through the bottom port.  Analogous to the top port, we have $s\,e^{i\psi} T_{21}\vec{\gamma}$ at $S_{mn}$ ($\vec{\gamma} = \hat{e}_{N-n+1}$), and $s\,e^{i\psi} T_{21}\vec{c}$ at the output ($\vec{c} = U_{\rm post} \vec{\gamma}$).
\end{enumerate}
Summing these terms, the output from port $m$ is:
\beq
	\vec{u}_m = \vec{a} + s\,e^{i\psi}(T_{11}\vec{b} + T_{21}\vec{c})
\eeq
The goal is to configure the MZI so that this output best approximates $\hat{u}_m$, the $m^{\rm th}$ column of target matrix $\hat{U}$.  This is done by minimizing the $L_2$ norm $\lVert \vec{u}_m - \hat{u}_m \rVert$.  Since $T$ and $U_{\rm post}$ are unitary, we have $\vec{a} \perp \vec{b} \perp \vec{c}$ and $|T_{11}|^2 + |T_{21}|^2 = 1$; applying these relations we find:
\begin{align}
	& \lVert \vec{u}_m - \hat{u}_m \rVert^2 = \underbrace{\lVert \hat{u}_m - \vec{a} \rVert^2 + |s|^2}_{\rm const} \nonumber \\
	& \qquad\quad - 2\,\text{Re}\bigl[s\,e^{i\psi}\bigl(T_{11}\langle\hat{u}_m|\vec{b}\rangle + T_{21} \langle\hat{u}_m|\vec{c}\rangle\bigr)\bigr]
\end{align}
The first two terms drop out as constants since they do not depend on the optimization variables $\theta_{mn}, \phi_{mn}$ (which determine $(T_{11},T_{21})$).  Since each MZI is preceded by a phase shifter, the phase of $\psi$ is also freely tunable; we therefore wish to perform the following maximization:
\begin{align}
	& \text{max}_{\theta, \phi} \text{max}_\psi \text{Re}\bigl[s\,e^{i\psi}\bigl(T_{11}\langle\hat{u}_m|\vec{b}\rangle + T_{21} \langle\hat{u}_m|\vec{c}\rangle\bigr)\bigr] \nonumber \\
	& \quad \propto \text{max}_{\theta, \phi} \bigl|T_{11} \langle\hat{u}_m|\vec{b}\rangle + T_{21} \langle\hat{u}_m|\vec{c}\rangle\bigr| \label{eq:max1}
\end{align}
subject to the constraint $|T_{11}|^2 + |T_{21}|^2 = 1$.  This is just optimizing a dot product, which amounts to setting the amplitude ratio:
\beq
	\frac{T_{11}}{T_{21}} = \frac{\langle \vec{b}|\hat{u}_m\rangle}{\langle \vec{c}|\hat{u}_m\rangle}
\eeq
We cannot measure $T_{11}, T_{21}, \vec{b}$, or $\vec{c}$ directly in an experiment.  Instead, we proceed as follows: first sweep the value of $\psi$ to obtain $\vec{a}$, which is the value of $u_i$ averaged over opposite phases $\psi$:
\beq
	\vec{a} = \frac{\vec{u}_m(\psi=0) + \vec{u}_m(\psi=\pi)}{2} = \frac{1}{2\pi}\int{\vec{u}_m(\psi)\d\psi}
\eeq
Next, once $\vec{a}$ is found, set $(\theta, \phi)$ to maximize the quantity:
\beq
	\text{max}_{\theta, \phi} \bigl|\langle \hat{u}_m | \vec{u}_m(\theta, \phi) - \vec{a} \rangle \bigr|
\eeq
which is mathematically the same as Eq.~(\ref{eq:max1}) and independent of $\psi$, but only relies on external output measurements.

The Ratio Method is closely related to mesh configuration methods based on matrix diagonalization \cite{Clements2016, Miller2013a, Miller2013b, RyanPaper2}.


\begin{figure}[tb]
\begin{center}
\includegraphics[width=1.00\columnwidth]{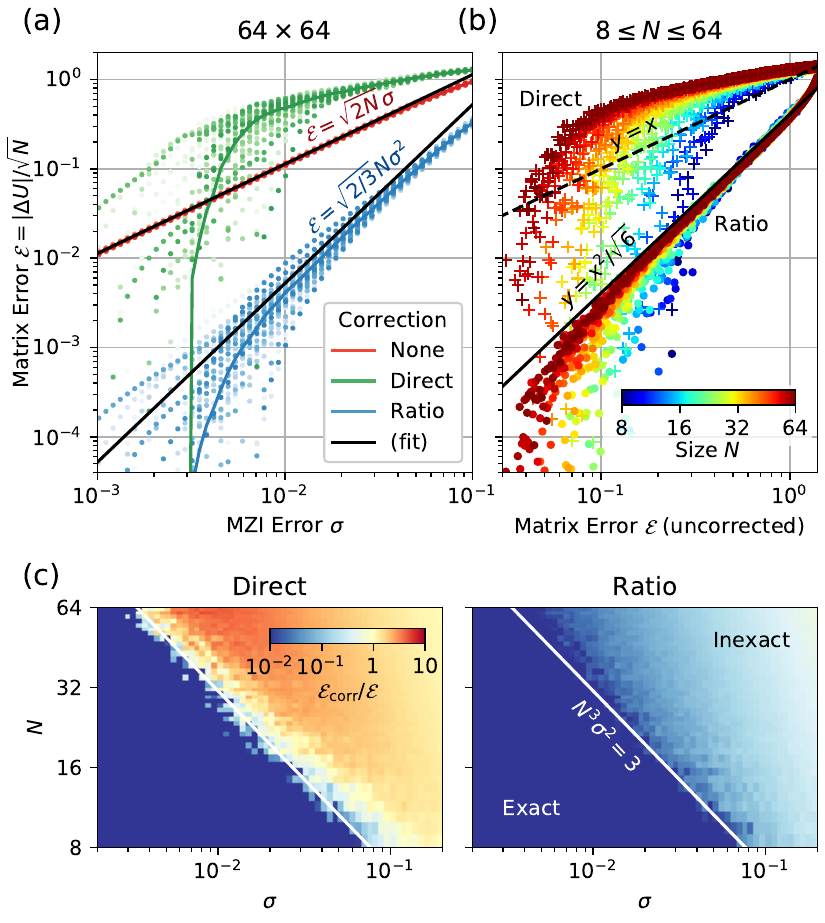}  
\caption{Performance of mesh calibration methods, \textsc{Reck} triangle.  (a) Matrix error $\mathcal{E}$ vs.\ MZI error $\sigma$ for $N=64$.  Dots are individual instances; the line traces the median.  (b) Uncorrected vs.\ corrected error, $8 \leq N \leq 64$.  (c) Correction factor $\mathcal{E}_{\rm corr}/\mathcal{E}$ as a function of $\sigma$ and $N$, showing the sharp transition for the direct method.  Data for (b-c) show medians for each $(N, \sigma)$ parameter pair.}
\label{fig:f5}
\end{center}
\end{figure}

\section{Performance Comparison}
\label{sec:perf}

To compare the strategies, we simulate the self-calibration of \textsc{Reck} meshes in the presence of component errors.  Here, the target unitaries $\hat{U}$ are sampled from the Haar measure, with random, Gaussian-distributed errors in the beamsplitter angles (i.e. $\langle \alpha\rangle_{\rm rms} = \langle \beta\rangle_{\rm rms} = \sigma$).  We consider mesh sizes in the range $8 \leq N \leq 64$ to analyze the scaling of the algorithms with mesh size.  Along the lines of Sec.~\ref{sec:errcorr}, we expect that error correction should allow perfect configuration when errors are low enough (coverage is order unity), and an error reduction of $\mathcal{E}_{\rm corr} \propto \mathcal{E}^2$ in the uncorrectable case.

Fig.~\ref{fig:f5}(a) shows the scaling of error metric $\mathcal{E}$ with $\sigma$ for a $64\times 64$ \textsc{Reck} mesh.  As expected, the uncorrected error increases linearly with $\sigma$, following Eq.~(\ref{eq:e0}).  For sufficiently small $\sigma$, the corrected error diverges to zero for both methods.  However, the direct method suffers a hard performance cutoff around $\sigma = 0.005$.  For realistic $\sigma \gtrsim 0.01$, the direct method actually performs worse than no correction at all!  In contrast, the ratio method performs well at both small and large $\sigma$.  Above the cutoff, it roughly follows the trend $\mathcal{E} = \sqrt{2/3}\,N\sigma^2$, the same relation derived for the local scheme in Eq.~(\ref{eq:ecorr}).  Unlike the local scheme, the ratio method requires no prior knowledge of the MZI imperfections, and can be configured using only output detectors.

By combining the two fits in Fig.~\ref{fig:f5}(a), we arrive at the following expression:
\beq
	\mathcal{E}_{\rm corr} = \frac{1}{\sqrt{6}}\mathcal{E}^2 \label{eq:ee}
\eeq
This relation is independent of $N$, and can be used to test the scaling of the algorithms as the matrix dimension increases.  Fig.~\ref{fig:f5}(b) plots $\mathcal{E}_{\rm corr}$ against $\mathcal{E}$ for $8 \leq N \leq 64$.  Predictably, there is a sharp drop for both schemes corresponding to perfect correction, but the threshold for such perfect correction decreases with $N$.  This is a serious challenge for the direct method, since $\mathcal{E}_{\rm corr} > \mathcal{E}$ in the imperfect correction regime.  On the contrary, the ratio method shows an improvement for the whole range of $N$, with the data asymptoting to Eq.~(\ref{eq:ee}), suggesting that the approach is scalable.

Fig.~\ref{fig:f5}(c) plots the same data in $(N, \sigma)$ space.  For both methods, we observe a transition when the coverage of the mesh drops below unity.  Since $\text{cov}(N) \sim e^{-N^3\sigma^2/3}$, this transition occurs roughly at $N^3\sigma^2 = 3$ (white curve).  Both methods work in the exact regime, but only the ratio method is successful when errors are large enough that $U$ cannot be represented exactly.

\begin{figure}[t!]
\begin{center}
\includegraphics[width=1.00\columnwidth]{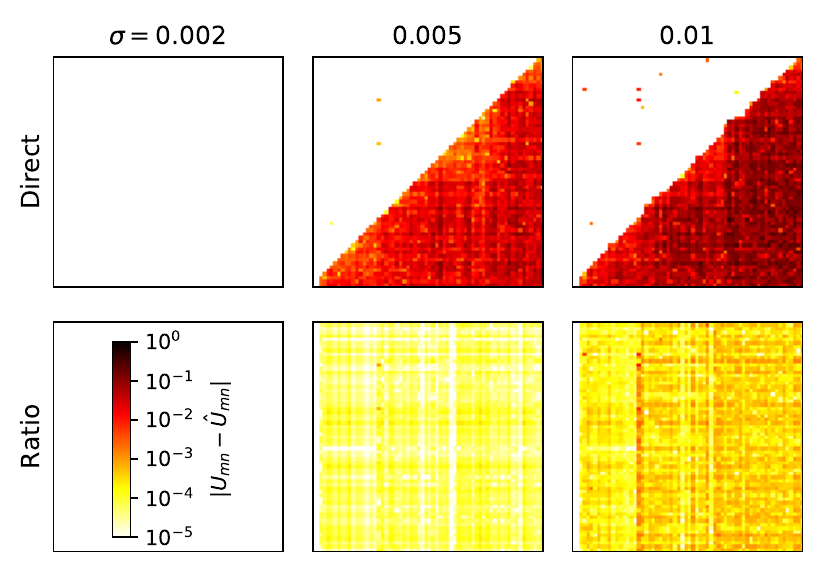}  
\caption{Matrix element errors $|U_{mn} - \hat{U}_{mn}|$ for $64\times 64$ \textsc{Reck} mesh as a function of MZI error $\sigma$ and calibration method.}
\label{fig:f6}
\end{center}
\end{figure}

Why does the direct method perform poorly in the uncorrectable error regime?  The structure of the error matrix $|U - \hat{U}|$  (Fig.~\ref{fig:f6}) sheds light on the problem.  The direct method only guarantees $U_{mn} = \hat{U}_{mn}$ for the upper-left triangle of entries $m + n < N$.  If these are exactly satisfied, the matrices will be equal.  However, even small errors are pushed to the lower-right triangle, where they cascade as the mesh is configured column by column.  This instability leads, in general, to a matrix that is only well configured for at most half of its entries.

\begin{figure}[t!]
\begin{center}
\includegraphics[width=1.0\columnwidth]{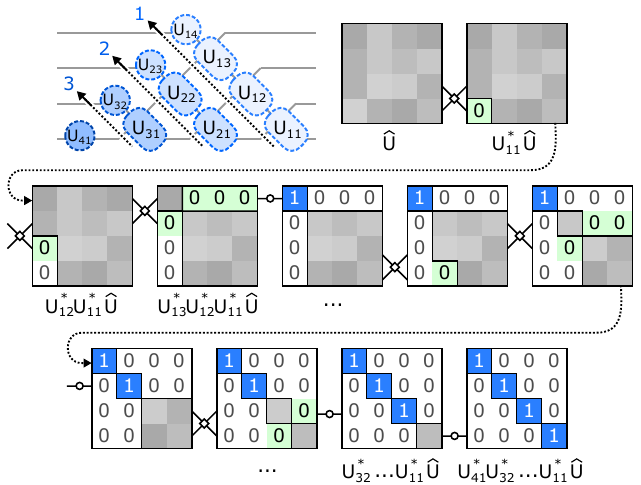}
\caption{Configuration of MZI mesh corresponds to diagonalizing a unitary matrix with $2\times 2$ blocks.}
\label{fig:f7}
\end{center}
\end{figure}

When following the ratio method, errors do not build up.  To understand the stability of the ratio method, it is helpful to relate the method to the $2\times 2$ block decomposition of a unitary matrix \cite{Reck1994} (Fig.~\ref{fig:f7}).  Given a target matrix $\hat{U}$, we wish to find $2\times 2$ blocks $U_{mn}$ such that $U_{N1}^\dagger\ldots U_{12}^\dagger U_{11}^\dagger \hat{U} = I$.  We start by configuring block $U_{11}$, which mixes the last two rows to zero out the lower-right element of the matrix.  Next, $U_{12}$ is configured to zero the element directly above.  The procedure is repeated until all elements in the lower diagonal have been zeroed, at which point the matrix equals the identity.

Because of MZI errors, not all off-diagonal terms can be zeroed.  If a term cannot be zeroed, it leaves a residual term $(|\alpha_{mn}-\beta_{mn}| - \theta_{mn})$ below the diagonal, where $\theta_{mn}$ is the target splitting ratio for MZI $(m, n)$, which is unrealizable since $|\alpha_{mn}-\beta_{mn}| > \theta_{mn}$.  Let $V^{(mn)} = U_{mn}^\dagger\ldots U_{12}^\dagger U_{11}^\dagger \hat{U}$ be the matrix after configuring $U_{mn}$ and define
\beq
	\epsilon_{mn} = \sum_{i=1}^{m-1} \sum_{j=i+1}^N |V^{(mn)}_{ij}|^2 + \sum_{j=N-n}^N |V^{(mn)}_{mj}|^2
\eeq
which is the sum of squares of all elements $V^{(mn)}_{ij}$ in the zero region below the diagonal (white and green in Fig.~\ref{fig:f7}).  Each imperfect configuration step adds a new element to this region, incrementing $\epsilon_{mn}$ by $(|\alpha_{mn}-\beta_{mn}| - \theta_{mn})^2$, the norm of the new element added.  The existing matrix elements are mixed around, but the norm for them does not change because the mixing is unitary.  Therefore, errors do not grow in the ratio scheme; they just get mixed into other matrix elements.  This is the critical difference between the direct and ratio schemes.  The final matrix will be close to the identity and therefore takes the form $V^{(N1)} \approx I + i H$ for some Hermitian $H$.  Therefore, $\lVert U - \hat{U}\rVert^2 \approx 2 \epsilon_{N1}$, and we have:
\begin{align}
	& \mathcal{E}_{\rm corr}^2 = \frac{2\langle \epsilon_{N1}\rangle}{N} \nonumber \\
	& = \frac{2}{N} \sum_{mn} \langle\text{max}(|\alpha_{mn}-\beta_{mn}| - \theta_{mn}, 0)^2 \rangle = \frac{2N^2\sigma^4}{3}
\end{align}
which is the same result as for the local scheme Eq.~(\ref{eq:ecorr}).

\section{Rectangular Mesh}
\label{sec:clem}

\begin{figure}[b!]
\begin{center}
\includegraphics[width=1.00\columnwidth]{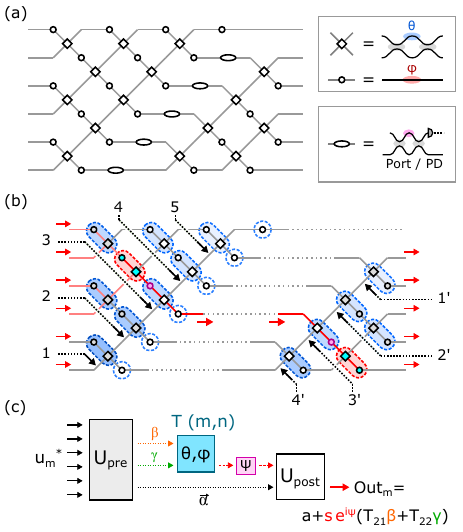}
\caption{Self-configuration of rectangular mesh.  (a) Modified \textsc{Clements} mesh with a row of drop ports and detectors along the diagonal.  (b) \textsc{Clements} mesh divided along the diagonal, showing the order or MZI configuration each triangle, according to the ratio method (reciprocal variant used for the first triangle).  (c) Flow of light when configuring an MZI in the reciprocal ratio method.}
\label{fig:f8}
\end{center}
\end{figure}

Compared to the \textsc{Reck} triangle, the rectangular \textsc{Clements} mesh has the advantages of increased compactness, reduced circuit depth, and relative insensitivity to fixed component losses \cite{Clements2016}.  However, it cannot be written as a cascade of diagonals, so the self-configuration techniques presented above cannot be used.  However, a simple modification to \textsc{Clements}---placing a diagonal of tunable drop ports and detectors---suffices to make the mesh programmable (Fig.~\ref{fig:f8}(a)).  The diagonal elements effectively split the mesh into two triangles, each which can be programmed independently.

First, following the \textsc{Clements} representation of an ideal MZI mesh \cite{Clements2016}, the target matrix is decomposed into two components $\hat{U} = \hat{U}_2\hat{U_1}$ for the left and right triangles.  Next, the diagonal ports are set to the ``cross'' state to collect all of the light along the diagonal.  This allows the left triangle to be programmed to $\hat{U}_1$ (Fig.~\ref{fig:f8}(b)), achieved by a reciprocal form of the ratio method described below.  Finally, the diagonal switches are used to isolate the inputs of the right triangle; this allows it to be programmed to $\hat{U}_2$ (up to an input phase) by the conventional ratio method (Sec.~\ref{sec:ratio}).

In the reciprocal form of the ratio method, instead of sending light into a single port and matching the output vector to a column of $\hat{U}$, we send in a column $u_m^*$ of $\hat{U}^\dagger$ as input and try to direct all the power to a single output.  This is analogous to the Reverse Local Light Interference Method (RELLIM) \cite{Miller2017}, but does not require internal detectors.  The MZIs are programmed along falling diagonals, but compared to Sec.~\ref{sec:ratio}, the order is reversed (bottom to top, down each diagonal).

\begin{figure}[b!]
\begin{center}
\includegraphics[width=1.0\columnwidth]{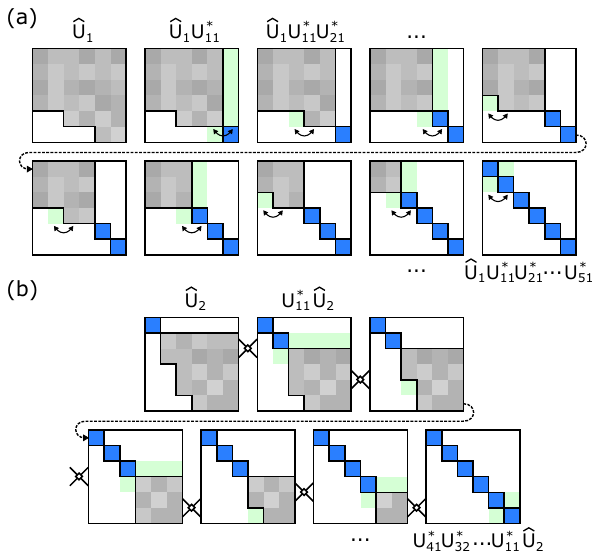}
\caption{Zeroing matrix elements when configuring the triangular meshes in Fig.~\ref{fig:f8}(b) to match $\hat{U}_1$ and $\hat{U}_2$.  (a) Reciprocal ratio method configures upper-left triangle to $\hat{U}_1$.  (b) Ratio method configures lower-right triangle to $\hat{U}_2$.}
\label{fig:f9}
\end{center}
\end{figure}

\begin{figure*}[tbp]
\begin{center}
\includegraphics[width=1.00\textwidth]{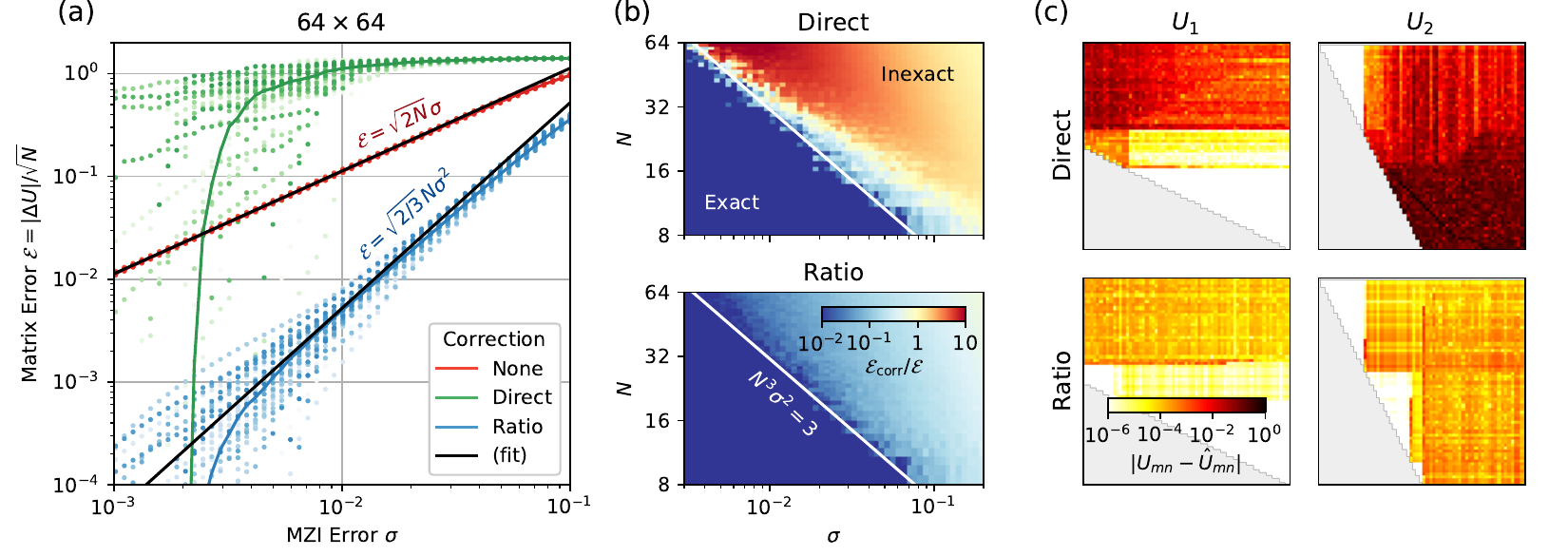}  
\caption{Accuracy of self-configured \textsc{Clements} mesh.  (a) Matrix error $\mathcal{E}$ vs.\ MZI error $\sigma$ for $N = 64$.  (b) Correction factor $\mathcal{E}_{\rm corr}/\mathcal{E}$.  (c) Matrix element errors $|U_{mn} - \hat{U}_{mn}|$ for $U_1$ and $U_2$ employing the direct and ratio methods ($N = 64, \sigma = 0.01$).}
\label{fig:f10}
\end{center}
\end{figure*}

When configuring MZI $(m, n)$, all upstream components have been configured, while downstream components have not.  The input-output relation is the product $v = U_{\rm post} T U_{\rm pre} \hat{u}_m^*$, where $U_{\rm pre}$ has been configured but $U_{\rm post}$ has not.  The light at output $m$ can take one of three paths: (1) bypassing the MZI, (2) entering the top port of the MZI, or (3) entering the bottom port of the MZI (Fig.~\ref{fig:f8}(c)), leading to a sum:
\beq
	v_m = a + s\, e^{i\psi} (T_{21}\beta + T_{22}\gamma)
\eeq
The MZI must be configured to direct all its output power to the bottom port ($T_{21}/T_{22} = (\beta/\gamma)^*$).  In RELLIM \cite{Miller2017}, this is accomplished with the use of internal detectors.  However, external detectors can also be used, even if the downstream MZIs have not been calibrated to a ``cross'' state.  As in Sec.~\ref{sec:ratio}, first the phase $\psi$ is swept and the value of $a$ is extracted from the average:
\beq
	a = \frac{v_m(\psi=0) + v_m(\psi=\pi)}{2} = \frac{1}{2\pi} \int{v_m(\psi)\d\psi}
\eeq
Next, the phases of the MZI are set to maximize:
\beq
	\text{max}_{\theta,\phi} | v_m - a |
\eeq
As with the \textsc{Reck} scheme, we can prove that this configuration method is resistant to errors by visualizing how each configuration step zeroes out entries in the target matrix.  Each matrix has $\approx \tfrac{3}{4}N^2$ free entries and $\approx \tfrac{1}{4}N^2$ zeroes below the diagonal.  Each mesh is self-configured to eliminate the remaining nonzero elements in the lower triangle, which takes $\approx \tfrac{1}{4}N^2$ steps, half the number of steps as the \textsc{Reck} triangle.  In each step in the reciprocal ratio method (Fig.~\ref{fig:f9}(a)), the target matrix is {\it right}-multiplied by a $2\times 2$ block to zero out a matrix entry; after $\approx \tfrac{1}{4}N^2$ steps, the upper triangle of Fig.~\ref{fig:f8}(b) is configured and $\hat{U}_1$ has been diagonalized.  Likewise, the conventional ratio method configures the lower triangle and diagonalizes $\hat{U}_2$ (Fig.~\ref{fig:f9}(b)).  The remaining phase shifts along the diagonal can be set by inspection.

The accuracy of the self-configured meshes is plotted in Fig.~\ref{fig:f10}.  Again, we see that the ratio method successfully corrects MZI errors and follows the same relation $\mathcal{E} = \sqrt{2/3}\,N\sigma^2$ observed for the \textsc{Reck} triangle (Fig.~\ref{fig:f10}(a)).  The direct method can also be used to configure the sub-triangles in Fig.~\ref{fig:f8}(b), but shows poor performance in the large-$\sigma$ regime where MZI errors cannot be exactly corrected.  The boundary between the exact and inexact correction regime is the same (Fig.~\ref{fig:f10}(b)) owing to the fact that the MZI splitting angles in \textsc{Reck} and \textsc{Clements} meshes have the same statistics over the Haar measure (Sec.~\ref{sec:stats}).  Visualizing the matrix imperfections (Fig.~\ref{fig:f10}(c)) again reveals the harmful cascading of errors in the direct method, which cause certain parts of the matrix to be well-configured while other parts are not.  This cascading effect is avoided in the ratio method for the same reasons applicable to \textsc{Reck} (Sec.~\ref{sec:perf}).

\section{Discussion}

Component imprecision will limit practical performance as multiport interferometers grow larger---barring a breakthrough in fabrication accuracy, some form of self-calibration or error correction will need to be employed.  In this paper, we have presented a method based on power ratios that can operate without any internal detectors or knowledge of the component imperfections.  This algorithm is applicable to any triangular mesh, but can be extended to rectangular meshes by adding a single diagonal of drop ports, a small amount of additional complexity as the mesh size grows large.  The accuracy of our algorithm is guaranteed by the algorithmic stability of unitary matrix diagonalization, and follows the asymptotic form $\mathcal{E}_{\rm corr} \propto N\sigma^2$ over the Haar measure with independent Gaussian component errors.  Employing this algorithm suppresses matrix errors by a quadratic factor: $\mathcal{E}_{\rm corr} = \mathcal{E}^2/\sqrt{6}$, allowing MZI meshes to scale to large sizes ($N > 64$) without unreasonable demands on fabrication tolerance.

Two limitations to our algorithm merit future work.  First, it relies heavily on unitarity and will fail to correct non-unitary errors.  Non-unitary mesh architectures and calibration are important topics for future study, as many ultra-compact or energy-efficient component designs \cite{Harris2014, Feldmann2019, Leinse2005, Haffner2019} involve some loss (often phase-dependent).  Second, the algorithm will only be effective if uncorrected errors are reasonably small to begin with.  For large enough meshes ($N \gtrsim 1/\sigma^2$), $\mathcal{E} \sim 1$ and errors will be uncorrectable.  This may lead to a fundamental limit on size, closely tied to the transition between ballistic and diffusive transport of light \cite{Pai2019}.  For such extreme sizes, entirely new crossing designs \cite{Miller2015} or mesh architectures \cite{Polcari2018, Lopez2019, Fldzhyan2020, Saygin2020, Tanomura2019} may be required.

\section*{Acknowledgements}

S.B.\ is supported by a National Science Foundation (NSF) Graduate Research Fellowship (grant no. 1745302).  D.E.\ acknowledges funding from the Air Force Office of Scientific Research (AFOSR) (grant no. FA9550-20-1-0113 and FA9550-16-1-0391).

R.H., S.B., and D.E. are inventors on patent application No.~96/196,301 assigned to MIT and NTT Research that covers techniques to suppress component errors in interferometer meshes.

Source code implementing the algorithms described in this paper is available in the \textsc{Meshes} package \cite{Meshes}. Scripts to generate the paper figures are provided in the Supplementary Material \cite{Supp}.


\appendix

\section{Correlated Errors}
\label{sec:appcorr}

In this paper, we have assumed an uncorrelated error model, where $\alpha_n$ and $\beta_n$ are independent random variables sampled from a zero-mean Gaussian with standard deviation $\sigma$.  In practice, the fabrication imperfections that lead to splitter errors have a long correlation length, so errors will be strongly correlated.  In this section, we show that:
\begin{enumerate}
	\item Averaged over the Haar measure, inter-MZI correlations cancel out.  Therefore, only correlations within each MZI (i.e.\ between $\alpha_n$ and $\beta_n$) need be considered.
	\item For most unitaries, the effect of the symmetric error $\alpha_n + \beta_n$ is dominant.
\end{enumerate}
The error metric Eq.~(\ref{eq:err}) can be expanded to second order, accounting for all correlations:
\begin{align}
	& \mathcal{E}^2 = \frac{1}{N} \sum_{mn} \Bigl(
		\Bigl\langle\frac{\partial U}{\partial\alpha_m}\Bigr|\frac{\partial U}{\partial\alpha_n}\Bigr\rangle 
		\langle \alpha_m \alpha_n\rangle + 
		\Bigl\langle\frac{\partial U}{\partial\alpha_m}\Bigr|\frac{\partial U}{\partial\beta_n}\Bigr\rangle 
		\langle \alpha_m \beta_n\rangle\nonumber \\
		& \qquad\quad +\,\Bigl\langle\frac{\partial U}{\partial\beta_m}\Bigr|\frac{\partial U}{\partial\alpha_n}\Bigr\rangle 
		\langle \beta_m \alpha_n\rangle + 
		\Bigl\langle\frac{\partial U}{\partial\beta_m}\Bigr|\frac{\partial U}{\partial\beta_n}\Bigr\rangle 
		\langle \beta_m \beta_n\rangle \Bigr) \label{eq:corr}
\end{align}
with the matrix inner product $\langle V | W \rangle = \text{tr}(V^\dagger W)$.

\begin{figure}[b!]
\begin{center}
\includegraphics[width=1.00\columnwidth]{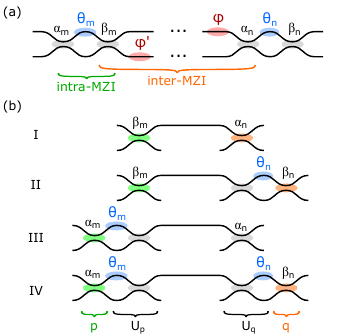}
\caption{(a) Intra- and inter-MZI beamsplitter error correlations.  (b) Four inter-MZI correlation types.}
\label{fig:f11}
\end{center}
\end{figure}

Correlations can be classified into two types: intra-MZI and inter-MZI (Fig.~\ref{fig:f11}(a)).  We first show that, {\it averaged over the Haar measure}, inter-MZI correlations are zero or at least very small.  Consider an arbitrary inter-MZI pair of splitters $(p \in \{\alpha_m, \beta_m\}, q \in \{\alpha_n, \beta_n\})$.  The unitary takes the form:
\begin{align}
	U &= U_{\rm post} S(\tfrac\pi4 \!+\! q) U_q\!
	\begin{bmatrix} e^{i\phi} & 0 \\ 0 & 1\end{bmatrix}
	\!U_{\rm int}\!
	\begin{bmatrix} 1 & 0 \\ 0 & e^{i\phi'}\end{bmatrix} 
	\!U_p S(\tfrac\pi4 \!+\! p)
	U_{\rm pre} \nonumber \\
	& \ \ \ \ + P(p) + Q(q) \label{eq:apq}
\end{align}
where $S$ is the symmetric splitter matrix (Eq.~(\ref{eq:tprod2}))
\bea
	S(\psi) & = & e^{i\psi\sigma_x} = \begin{bmatrix} \cos\psi & i\sin\psi \\ i\sin\psi & \cos\psi \end{bmatrix} \nonumber \\ 
	S'(\psi) & = & i S(\psi) \sigma_x = i\sigma_x S(\psi)
\eea
and $\sigma_x = [[0, 1], [1, 0]]$ is the Pauli matrix.

The terms $P(p)$ and $Q(q)$ in Eq.~(\ref{eq:apq}) correspond to paths of light that pass through at most one splitter.  These are mutually orthogonal and do not contribute to the correlation in Eq.~(\ref{eq:corr}).  Error correlations only arise from the first term, corresponding to paths of light that pass through both splitters $p$ and $q$.  The resulting inner product is independent of $U_{\rm pre}$ and $U_{\rm post}$ and takes the form (at $p = q = 0$):
\begin{align}
	\Bigl\langle\frac{\partial U}{\partial p}\Bigr|\frac{\partial U}{\partial q}\Bigr\rangle
	& = \Bigl\langle 
		S(\tfrac\pi4) U_q\!
		\begin{bmatrix} e^{i\phi} & 0 \\ 0 & 1\end{bmatrix} \!U_{\rm int}\!
		\begin{bmatrix} 1 & 0 \\ 0 & e^{i\phi'}\end{bmatrix} \!U_p
		S(\tfrac\pi4) 
		\sigma_x \Bigr| \nonumber \\
	& \qquad \Bigl| 
		\sigma_x
		S(\tfrac\pi4) U_q\!
		\begin{bmatrix} e^{i\phi} & 0 \\ 0 & 1\end{bmatrix} \!U_{\rm int}\!
		\begin{bmatrix} 1 & 0 \\ 0 & e^{i\phi'}\end{bmatrix} \!U_p
		S(\tfrac\pi4) 
	\Bigr\rangle \label{eq:ip}
\end{align}
In general, this quantity is nonzero.  However, the phases $(\phi, \phi')$ are uniformly distributed over $[0, 2\pi)$ for Haar-random unitaries.  Therefore, in the ensemble average over the Haar measure, the phase-dependent terms in Eq.~(\ref{eq:ip}) cancel out.  This means that each path from splitter $p$ to $q$, which has a separate phase dependence, can be considered separately in the inner product.  Consider the path from output $i \in \{1, 2\}$ to input $j \in \{1, 2\}$ ($i = j = 1$ shown in Fig.~\ref{fig:f11}(b)).  Focusing on a single path, we can ignore ($\phi, \phi')$ and replace $U_{\rm int} \rightarrow \hat{e}_j \hat{e}_i^T$:
\begin{align}
	\Bigl\langle\frac{\partial U}{\partial p}\Bigr|\frac{\partial U}{\partial q}\Bigr\rangle_{ij}
	& \propto \bigl\langle S(\tfrac\pi4) U_q \hat{e}_j \hat{e}_i^T U_p S(\tfrac\pi4) \sigma_x \bigr| \nonumber \\
	& \qquad \bigl| \sigma_x S(\tfrac\pi4) U_q \hat{e}_j \hat{e}_i^T U_p S(\tfrac\pi4) \bigr\rangle \nonumber \\
	& = (\hat{e}_j^T U_q^\dagger\sigma U_q \hat{e}_j) (\hat{e}_i^T U_p\sigma U_p^\dagger \hat{e}_i) \label{eq:ip2}
\end{align}
There are four cases to be considered (Fig.~\ref{fig:f11}(b)).  If $p$ is directly adjacent to the connecting path (cases I-II), then $U_p$ is the identity and $\hat{e}_i^T U_p \sigma U_p^\dagger \hat{e}_i = \hat{e}_i^T \sigma \hat{e}_i = 0$.  Likewise, if $q$ is adjacent to the path (cases I and III), $\hat{e}_j^T U_q^\dagger\sigma U_q \hat{e}_j = 0$.  Only in case IV does Eq.~(\ref{eq:ip2}) lead to a nontrivial inner product:
\beq
	\Bigl\langle\frac{\partial U}{\partial p}\Bigr|\frac{\partial U}{\partial q}\Bigr\rangle_{ij}
	\propto \sin(\theta_m)\sin(\theta_n) \label{eq:ip3}
\eeq
This is very small for the majority of MZIs, where the splitter angles ($\theta_m$, $\theta_n$) cluster tightly around zero.  Moreover, while it is always possible to find a matrix decomposition with only positive $\theta$ (e.g.\ distribution of Eqs.~(\ref{eq:pph}-\ref{eq:pth})), one can also sample from the Haar measure employing both positive and negative $\theta$ with equal probability; in this case, under the ensemble average, Eq.~(\ref{eq:ip3}) vanishes and all inter-MZI correlations are zero.

On the other hand, correlations {\it within an MZI}, i.e.\ between $\alpha_n$ and $\beta_n$, always matter.  The matrix error of a single MZI is:
\bea
	\lVert \Delta U \rVert^2 & = & 2\bigl[\langle \alpha_n^2\rangle + \langle \beta_n^2\rangle + 2\cos(\theta_n) \langle \alpha_n\beta_n \rangle \bigr] \nonumber \\
	& = & 2\bigl[\cos^2(\theta_n/2) \langle (\alpha_n+\beta_n)^2\rangle \nonumber \\
	& & \ \ \ + \sin^2(\theta_n/2)\langle (\alpha_n-\beta_n)^2\rangle \bigr] 
\eea
For the large majority of MZIs, $\theta_n \approx 0$ and the symmetric error dominates $\lVert \Delta U\rVert^2$.  The normalized error for the whole mesh is approximately:
\beq
	\mathcal{E} \approx \sqrt{N} \langle(\alpha+\beta)\rangle_{\rm rms}
\eeq
which in the case of uncorrelated $(\alpha, \beta)$ reduces to the form derived in the main text: $\mathcal{E} = \sqrt{2N}\,\sigma$.

\begin{figure}[tbp]
\begin{center}
\includegraphics[width=1.00\columnwidth]{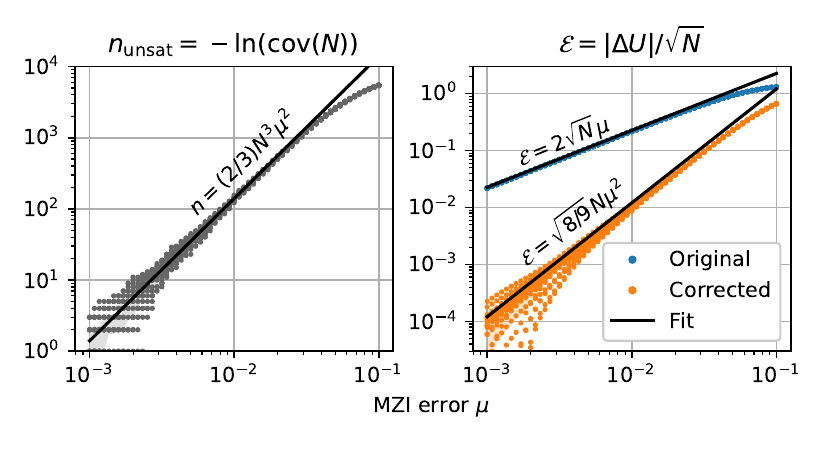}
\caption{Correlated errors: unsatisfied MZI fraction $n_{\rm sat}$ and matrix error $\mathcal{E}$ as a function of splitter error $\mu$.  $N = 128$ \textsc{Reck} mesh.}
\label{fig:f12}
\end{center}
\end{figure}

For completeness, we now consider the case of fully correlated splitter errors, i.e.\ $\alpha_n = \beta_n = \mu$, where $\mu$ is a constant.  Systematic effects such as imperfect coupler design or fabrication errors with long correlation length will lead to this situation, as will operating the mesh away from the coupler design wavelength.  The uncorrected error amplitude is (compare Eq.~(\ref{eq:e0})):
\beq
	\mathcal{E} = 2\sqrt{N} \mu \label{eq:e0mu}
\eeq
Following the analysis of Eq.~(\ref{eq:pph}-\ref{eq:cov}), the coverage of unitary space is found to be (compare Eq.~(\ref{eq:cov})):
\beq
	\text{cov}(N) = e^{-(2/3)N^3\mu^2}
\eeq
Finally, in the presence of error correction, the matrix error becomes (compare Eqs.~(\ref{eq:ecorr}, \ref{eq:ee})):
\beq
	\mathcal{E}_{\rm corr} = \sqrt{8/9} N\mu^2 = \frac{1}{\sqrt{18}} \mathcal{E}^2 \label{eq:ecmu}
\eeq
Eqs.~(\ref{eq:e0mu}-\ref{eq:ecmu}) are plotted against numerical data in Fig.~\ref{fig:f12}.

\section{Time and Computational Cost}

\begin{table*}[t!]
\begin{center}
\begin{tabular}{lc|cc|c|c|ccc}
\hline\hline
& & \multicolumn{2}{c|}{Mesh$^*$} & MVMs & CFLOPs & \multicolumn{3}{c}{Caveats$^\dagger$} \\
\hline
Progressive & \cite{Miller2013a, Miller2013b} & R & & $N^2$ & $N^3$ & & (D) \\
RELLIM & \cite{Miller2017, Pai2020} & R & C & $N$--$N^2$ & $N^3$ & & (D) \\
In-situ & \cite{Hughes2018} & R & C & $NT$ & $N^3 T$ & & (D) \\
\hline
SGD & \cite{Pai2019} & R & C & $BT$ & $N^2 BT$ & (C) & \\
Numerical & \cite{Mower2015, Burgwal2017} & R & C & $NT$ & $N^3 T$ & (C) & \\
Local & \cite{SaumilPaper} & R & C & -- & $N^2$ & (C) & \\
\hline
Direct & (this\ & R & & $N^2$ & $N^3$ & & & (S) \\
Ratio & \ work) & R & & $N^2$ & $N^3$ & & \\
\hline\hline
\end{tabular}
\caption{Comparison of a representative sample of optimization algorithms.  Scaling of computational cost plotted for in-situ resources (number of MVM calls), and in-silico resources (CFLOPs), where $N$ is the mesh size, $T$ is the number of optimization steps, and $B$ is the SGD batch size.  $^*$Mesh types: (R)eck, (C)lements.  $^\dagger$Caveats include (D) requires internal detectors, (C) requires pre-calibration of errors, (S) stability issues in presence of errors.}
\label{tab:ts2}
\end{center}
\end{table*}

Many programming algorithms for photonic meshes have been reported in the literature.  Table~\ref{tab:ts2} summarizes several leading approaches, evaluated in terms of their generality (applicable mesh types) and time / computational cost.  While this list does not claim to be exhaustive, it provides a representative sample and sheds light onto the limitations and tradeoffs of past approaches, where to date all optimization schemes require either (1) accurate pre-calibration of the hardware errors, or (2) $O(N^2)$ internal photodetectors used to monitor power at intermediate points on the mesh.  Unique among its peers, our algorithm lacks both requirements, allowing its use on uncalibrated ``zero-change'' photonic hardware and requiring coherent control / measurement only over inputs and outputs.

\begin{figure}[t!]
\begin{center}
\includegraphics[width=1.00\columnwidth]{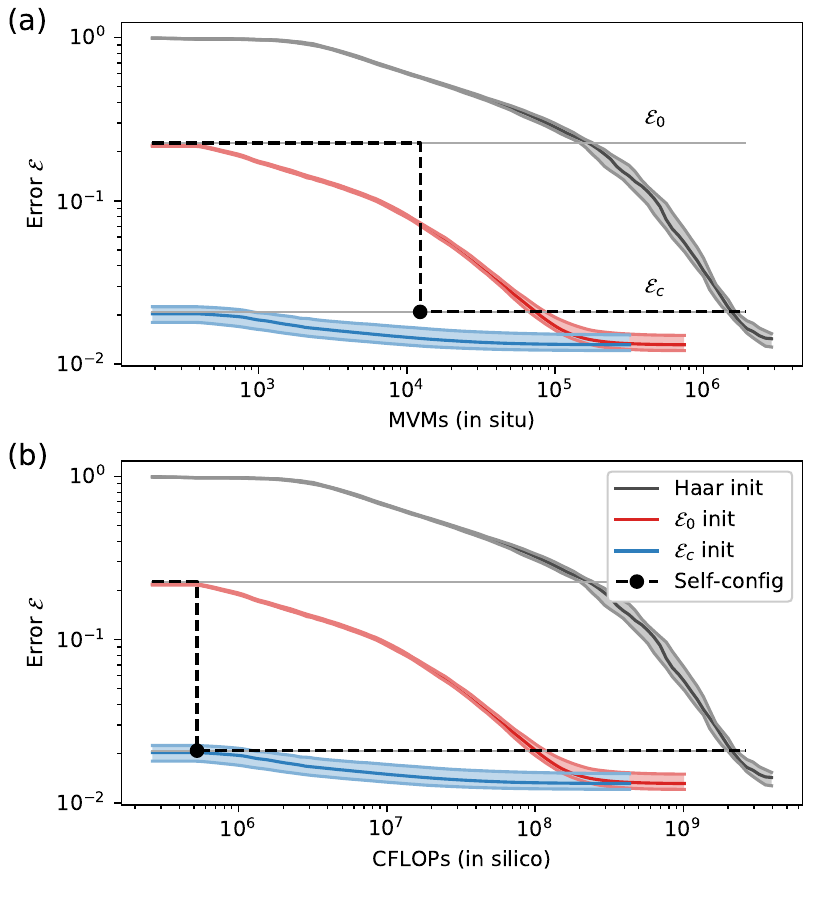}
\caption{Convergence of L-BFGS-B for an imperfect \textsc{Reck} mesh ($\sigma = 0.02$), minimizing the $L_2$ norm.  The convergence curve depends on the initial solution, and is compared with the result from self-configuration.  (a) Plot in terms of in-situ MVMs to solution.  (b) In-silico CFLOPs to solution.}
\label{fig:f13}
\end{center}
\end{figure}

Moreover, our algorithm is competitive in terms of computational resources.  The required resources of an algorithm depend on whether it is performed in situ (for correction of unknown errors) or in silico (for known, calibrated errors).  For in-situ algorithms, the number of matrix-vector multiplication (MVM) calls is most relevant.  The ratio method takes approximately $3N^2$ calls (2 to obtain vector $a$, 4 calls to optimize $(\theta, \phi)$ per MZI), which is comparable to progressive self-configuration and the initial RELLIM proposal \cite{Miller2017, Miller2013a, Miller2013b} (although RELLIM with internal detectors can run in parallel in $O(N)$ time \cite{Pai2020}).  The direct method can be performed with $2N^2$ calls since the vector $a$ does not need to be measured.  On the other hand, numerical optimization with $T$ time steps takes $3NT$ calls (three calls per step, first to estimate $U$, next to back-propagate $U-\hat{U}$, and finally one forward-pass step to compute the gradient with respect to phase shifts \cite{Hughes2018}).  Fig.~\ref{fig:f13}(a) shows the convergence of a $64\times 64$ \textsc{Reck} mesh in terms of MVMs under the L-BFGS-B algorithm.  Convergence depends strongly on the mesh initialization \cite{Pai2019}.  Initializing to Haar-random unitaries is a significant improvement over random phase shifts, since a mesh with random phase shifts has a banded structure that leads to vanishing gradients with respect to matrix elements far from the diagonal.  Even with Haar initialization, the algorithm takes thousands of steps and millions of MVMs to converge to the accuracy $\mathcal{E}_c$ reached by our algorithm.  Initializing to the phases of an ideal (error-free) \textsc{Reck} mesh helps considerably, but optimization still takes $5\times$ more calls.  Stochastic gradient descent (SGD) has been proposed as a solution to speed up the optimization, since a batch of $B < N$ columns is used rather than the whole matrix; however, one observes a tradeoff between batch size and  iteration count, so the overall resource requirement is higher \cite{Pai2019}.

Fig.~\ref{fig:f13}(b) shows the performance gap with respect to the in-silico case, where accurate calibration of the errors $(\alpha, \beta)$ allows the phases $(\theta, \phi)$ to be computed numerically.  Here, optimization protocols require $O(N^3)$ complex FLOPs (CFLOPs), equivalent to one matrix-matrix multiplication per time step, leading to a scaling of $N^3 T$ (SGD scales as $N^2 B T$).  The ratio method requires approximately $N^2/2$ Givens rotations to $U_{\rm post}$ with $4N$ CFLOPs each, for a total of $2N^3$ CFLOPs, and thus runs about $10^2\times$ faster (the direct method is similarly fast).  As before, numerical optimization can still be somewhat helpful as a means for further refinement of the solution.

However, for in silico optimization, the local correction method \cite{SaumilPaper} is superior, as it takes only $N^2$ CFLOPs and is parallelizable to $N$ time steps.


\bibliography{PaperRefs}{}
\bibliographystyle{IEEEtranN}

\end{document}